\documentclass[aps,pre,reprint,twocolumn,groupedaddress]{revtex4-2}

\newcommand{\bra}{\langle}
\newcommand{\ket}{\rangle}
\newcommand{\ve}{\varepsilon}
\newcommand{\be}{\begin{equation}}
\newcommand{\ee}{\end{equation}}
\newcommand\Sk{\mbox{\textit{Sk}}}
\newcommand\Pe{\mbox{\textit{Pe}}}

\newcommand{\lam}{\lambda}
\newcommand{\bwt}{\begin{widetext}}
\newcommand{\ewt}{\end{widetext}}

\usepackage{amssymb}
\usepackage{amsmath}
\usepackage{xcolor}
\usepackage{graphicx}
\usepackage[hidelinks]{hyperref} 
\begin{document}

\title{An advection-diffusion process with proportional resetting}

\author{J. Kevin Pierce}
\email[]{kevpierc@mail.ubc.ca}
\affiliation{Department of Geography, The University of British Columbia, Vancouver V6T 1Z4, Canada}

\date{\today}

\begin{abstract}
This paper presents a diffusion process with a novel resetting mechanism in which the amplitude of the process is instantaneously converted to a \textit{proportion} of its value at random times.
This model is described by a Langevin equation with both additive Gaussian white noise and multiplicative Poisson shot noise terms.
The distribution function obeys a pantograph equation, a functional partial differential equation evaluated at two amplitudes simultaneously.
From this equation the exact statistical moments and steady-state distribution of the process are calculated.
The distribution interpolates between exponential and Gaussian extremes depending on the proportion of the amplitude lost in each reset.
These results will be useful for applications in which stochastic quantities are suddenly reduced in proportion to their values due to random events.
\end{abstract}

\maketitle

\section{Introduction}

In recent times a number of works have evaluated the effects of stochastic resetting on diffusion processes.
Stochastic resetting shifts a quantity to a different value at random times \cite{Montero2017,Evans2020}.
Two key features can result.
First, resetting can \textit{stabilize} an otherwise unbounded diffusion process. 
In this instance, a non-equilibrium steady state results. This state is characterized by circulation of probability, rather than detailed balance \cite{Eule2016}.
Second, stochastic resetting can \textit{reduce} the mean first passage time of a stochastic process to a target.
Due to this second property, resetting can be used to optimize the performance of random searches \cite{Evans2011a,Evans2013}. 
Resetting therefore finds vast applications in machine learning, finance, data mining, foraging behavior, microbiology, and a number of other topics.

Besides a few antecedents (see \cite{Montero2017} for a historical review), stochastic resetting was largely established in the works of Evans, Manjumdar, and collaborators \cite{Evans2011,Evans2011a,Evans2013}.
These studies represent an underlying diffusion process $u(t)$ as a free Brownian motion. 
Resetting is introduced as the instantaneous transfer of the amplitude $u(t)$ of this diffusion process to a fixed reset value $u_r$. These resets occur randomly, at Poisson-distributed times.

Numerous generalizations of this process have been developed.
Examples include resetting to a previous maximum \cite{Majumdar2015}, resetting with memory of a former value \cite{Boyer2014}, resetting with state-dependent rates \cite{Roldan2017,Pinsky2020}, resetting with time-dependent rates \cite{Pal2016,Shkilev2017}, resetting with a delay upon arrival \cite{Maso-Puigdellosas2019}, resetting with a duration \cite{Bodrova2020,Pal2019}, resetting in a potential \cite{Pal2015,Pal2016}, resetting in a confined domain \cite{Chatterjee2018, Durang2020}, resetting with non-Poissonian times \cite{Eule2016,Nagar2016}, resetting of fluctuating interfaces \cite{Gupta2016,Durang2014}, resetting of fractional diffusion processes \cite{Sousa2018}, and resetting of finite-velocity diffusion processes \cite{Evans2018,Masoliver2019}. Several studies have considered resetting with many-particle interactions \cite{Basu2019, Magoni2020}. This is only a small sample of the generalizations available in the literature. The review articles \cite{Montero2017,Evans2020} provide a comprehensive account.

In this paper I would like to consider a new \textit{proportional resetting} process.
A proportional reset is considered to instantaneously convert a stochastic quantity $u$ to a proportion $\ve u$ of its value at random times, rather than to a fixed value $u_r$ or some other variant as above.
The \textit{resetting proportion} $\ve$ ranges over $0 < \ve < 1$. 
Proportional resetting is a new type of state-dependent resetting. The magnitude of each reset depends on the value of the process.

The objective of this paper is to characterize the proportional resetting process by (1) formulating its governing equations and (2) solving these equations for the statistical moments and steady state distribution.
The motivation is to provide a model which can be adapted to applications.
For example, in a granular gas, collisions convert the relative speed $u$ of two particles quasi-instantaneously to $\ve u$, $\ve$ being the coefficient of restitution \cite{Brilliantov2004}.
In a soil with some amount $u$ of a contaminant, rainfall events leach the contaminant out to an amount $\ve u$  \cite{Suweis2011,Suweis2010,Mau2014}.

Other possible contexts for proportional resetting include vegetation biomass with fires \cite{Clark1989, Odorico2006a}, population dynamics with environmental changes \cite{Wu2008a, Calabrese2017}, asset valuation with market crashes \cite{Merton1976,Kou2002}, and queueing with random clearing events \cite{Stidham1974,Kella2003}.
In all of these cases, we can imagine a stochastic quantity taking quasi-instantaneous losses due to random events, where the amount of each loss is some \textit{proportion} of the quantity present before the event
In this work I will investigate such a quantity, assuming that the underlying evolution in the absence of stochastic resetting is an advection-diffusion process.

The paper is outlined as follows. 
First, Sec. \ref{sec:methods} presents Langevin and forward Fokker-Planck equations for an advection-diffusion process with proportional resetting. Sec. \ref{sec:results} obtains the statistical moments and steady state distribution of this process. Here, limiting forms for the distribution are also derived for extremely weak and strong resets ($\ve \rightarrow 1$ and $\ve \rightarrow 0$ respectively).
Sec. \ref{sec:disc} discusses the key results, suggests next steps for research, and establishes the governing equation for the mean first passage time to the origin for the proprotional resetting process.
Finally, Sec. \ref{sec:conclusion} summarizes the main contributions of the paper.

\section{Stochastic evolution with proportional resetting}
\label{sec:methods}

\subsection{Construction of the Langevin equation}

 \begin{figure}
	\includegraphics{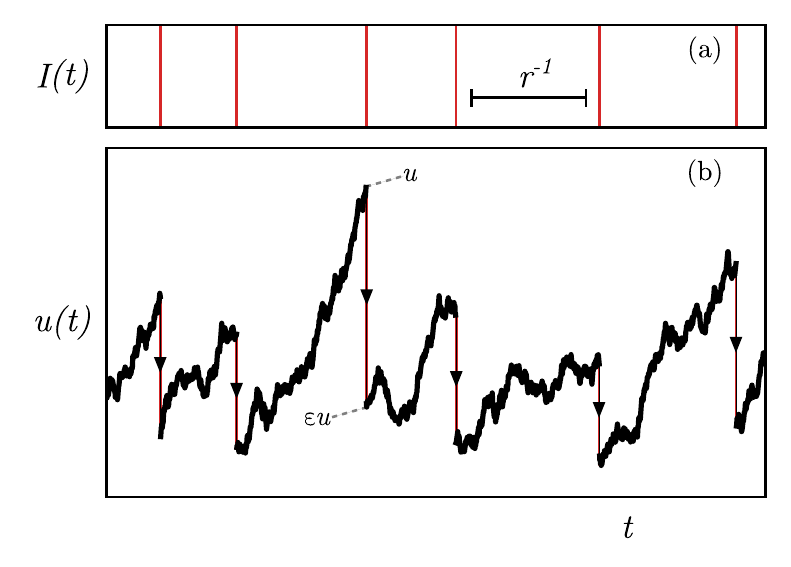}	
	\caption{The illustration shows a realization of the proportional resetting process.
		Resets occur at Poisson-distributed times with mean separation $1/r$ as shown in (a). Diffusion with resetting is displayed in (b). Resets appear as vertical red lines which carry the amplitude $u$ to a smaller value $\ve u$. \label{fig:fig1}} 	
\end{figure}

An advection-diffusion process with proportional resetting can be constructed with the Langevin equation
\begin{equation}
	\dot{u}(t) = \Gamma + \xi(t) - u(t) I(t). \label{eq:langevin}
\end{equation}
This equation describes the random evolution of the amplitude $u(t)$ subjected to three influence terms, one deterministic and two random.
$\Gamma$ is a constant advection term, while $\xi(t)$ is a Gaussian white noise having mean value $\bra \xi(t) \ket = 0$ and correlation function $\langle \xi(t)\xi(s) \rangle = 2 D \delta(t-s)$. $D$ is a diffusivity.

The quantity  $I(t)$ is a Poisson shot noise having rate $r$ and strength $1-\varepsilon$, given by:
\begin{eqnarray}
	I(t) = \sum_{k=0}^{N(t)} (1-\varepsilon)\delta(t-t_k).
\end{eqnarray}
This noise describes a sequence of $N(t)$ resets at times $t_1, \dots , t_{k},\dots, t_{N(t)}$. The time intervals between resets are considered as a stationary Poisson process with rate $r$, meaning inter-reset times $\tau = t_{k+1}-t_k$ are distributed exponentially: $P( \tau) = r e^{-r  \tau}.$ The mean time between resets is $1/r$. The number $N(t)$ of reset events in a time interval $t$ consequently follows a Poisson distribution: $P(N,t) = (rt)^N\exp(-r t)/\Gamma(N+1)$ \cite{Cox1965}.

Mathematically, (\ref{eq:langevin}) represents a jump-diffusion process with multiplicative Poisson noise \citep{Daly2006,Daly2010,Dubkov2016,Denisov2009}. 
According to (\ref{eq:langevin}), each reset converts $ u$ to $\ve u$. Thus $\varepsilon$ controls the amount by which resets affect $u$. The case $\ve \approx 1$ represents \textit{weak} resets which change $u$ very little, while $\ve\approx 0$ represents \textit{strong} resets.
Figure \ref{fig:fig1} illustrates a realization of the proportional resetting process. The unique feature is that the magnitude of each reset depends on the value of the process just before the reset occurred.

\subsection{Derivation of the forward Fokker-Planck equation}

The governing equation for the distribution of the advection-diffusion process with proportional resetting can be derived using the Markovian property of the process. Let $P(u,t|u_0,t_0)$ represent the distribution of $u(t)$ given that it started at $u_0$ and time $t_0$. $P(u,t|u_0,0)=P(u,t|u_0)$ provides a convenient abbreviation with $t_0=0$.
Because (\ref{eq:langevin}) is without memory, the process $u(t)$ is Markovian, and we can subdivide its density across an intermediate time with the Kolmogorov equation \cite{Gardiner1983}:
\be
	P(u,t+h|u_0) = \int_{-\infty}^\infty du' P(u,t+h|u',t)P(u',t|u_0). \label{eq:kolmo}
\ee
The timescale $h$ is considered very small. The Langevin equation (\ref{eq:langevin}) can be integrated across this timestep to evaluate the transition probability $P(u,t+h|u',t)$.

A complication arises because the Poisson noise is multiplicative. As a result, integration involves the prescription dilemma of stochastic calculus \citep{Risken1989,Gardiner1983,Suweis2011}. Here the It\^{o} interpretation (lower endpoint integration) is the physical choice, because reset magnitudes depend exclusively on the value of $u$ before reset.
With this interpretation, $P(u,t+h|u',t)$ derives to first order in $h$ as
\begin{multline}
	P(u,t+h|u',t)= r h \delta\big( u - \varepsilon u'\big) \\+ \frac{1-rh}{2}\sum_{s=\pm 1}\delta\big(u - \big[u' + \Gamma h + s(2 Dh)^{1/2}\big]\big).
\end{multline}
The sign of the drift accumulated by the Gaussian white noise is $s = \pm 1$. This has equal probability ($1/2$) to be positive or negative.

The terms in this equation either involve $r h$, representing the probability that one reset occurred in the small time interval $h$, or $1-rh$, representing the probability no reset occurred.
Expanding the above equality to first order in $h$, inserting this into (\ref{eq:kolmo}), and taking $h \rightarrow 0$ produces
\begin{align}
	\begin{split}
	\dot{P}(u,t|u_0) = \hat{L} P(u,t|u_0)\\ - r P(u,t|&u_0) + \frac{r}{\varepsilon} P\Big(\frac{u}{\varepsilon},t|u_0\Big). \label{eq:fp}
	\end{split}
\end{align}
This is the forward Fokker-Planck equation for the advection-diffusion process under proportional resetting. 
The operator $\hat{L}$ is defined by $\hat{L}= -\Gamma \partial/\partial u + D \partial^2/\partial u^2,$ and $\dot{P} = \partial P/\partial t$.

Terms in (\ref{eq:fp}) with $\Gamma$ and $D$ represent advection and diffusion respectively. Terms involving $r$ represent proportional resetting.
These include gain from $u/\ve$ and loss from $u$.
The gain term transfers probability from a higher pre-reset value $u/\ve$ to the amplitude $u$ involved in the rest of the equation, while the loss term transfers probability from $u$ to a lower value $\ve u$.
Since (\ref{eq:fp}) involves two different scales of $u$, it is a \textit{functional} partial differential equation. Both $u$ and $u/\varepsilon$ are arguments of $P$.

\section{Solution of proportional resetting}
\label{sec:results}

\subsection{Calculation of the moments}

All of the time-dependent moments derive by multiplying (\ref{eq:fp}) by $u^l$, $l = 0,1,\dots$ and integrating out $u$.
This produces the recursive moment equations
\begin{equation}
\Big[\frac{d}{dt} + (1- \varepsilon^l ) \Big] \langle u ^l \rangle = Dl(l-1)\langle u^{l-2}\rangle + \Gamma l \langle u^{l-1} \rangle. \label{eq:moments}
\end{equation}
These equations are to be interpreted with $\langle u^l\rangle=0$ whenever $l<0$.
Solving the mean equation with initial condition $\langle u(0)\rangle = u_0$ provides
\begin{equation}
	\langle u(t) \rangle = \frac{\Gamma}{r(1-\varepsilon)}\Big[1-e^{-(1-\varepsilon)rt}\Big] + u_0 e^{-(1-\varepsilon)rt}.
\end{equation}
The initial value $u_0$ is apparently remembered for a time $\tau_1 = [(1-\ve) r]^{-1}$, after which the mean trends to the steady-state value
\be 
\bra u \ket = \frac{\Gamma}{r(1-\ve)}. \label{eq:mean}
\ee
One can calculate the higher moments in the same way, solving (\ref{eq:moments}) recursively for higher $l$. Inspection of (\ref{eq:moments}) indicates that the timescale $\tau_l = [(1-\ve^l)r]^{-1}$ controls the slowest decaying transient in the $l$-th order moment. The transient terms of higher moments decay progressively more slowly.
The longest timescale affecting the adjustment to steady state is $\tau_\infty = r^{-1}$, the average time between resets.

In steady state ($\partial P/\partial t = 0$), one finds second and third order moments
\begin{align}
	\langle u^2 \rangle &= \frac{2Dr + \Gamma^2}{r^2(1-\ve^2)}, \label{eq:second} \\
	\langle u^3 \rangle &=  \frac{12 D \Gamma}{r^2(1-\ve^3)(1-\ve)} + \frac{6\Gamma^3}{r^3(1-\ve^3)(1-\ve^2)(1-\ve)}. 	
\end{align}
Formulas (\ref{eq:mean}-\ref{eq:second}) provide the variance, $\sigma_u^2 = \bra u^2 \ket-\bra u\ket^2$:
\be \sigma_u^2 = \frac{2 D r + \Gamma^2}{r^2(1+\ve)(1-\ve)} \label{eq:var}. \ee
This result indicates that both diffusion ($D$) and advection ($\Gamma$) contribute to the spreading of the amplitude through time. This contrasts with the expectation $\sigma_u^2 = 2 D t$ of the problem without resetting, where the spreading is indifferent to advection. With proportional resetting there is a feedback between the magnitude of $\Gamma$ and the magnitude of typical resets: the larger the advection term, the larger the typical reset. Accordingly, the variance grows with $\Gamma$.

The Fisher skew coefficient $\Sk$ takes the value $\Sk = 0$ for a Gaussian distribution and $\Sk = 2$ for a one-sided exponential distribution. This is defined as 
\be \Sk = \frac{\bra u^3\ket - 3 \sigma_u^2 \bra u \ket - \bra u \ket^3}{\sigma_u^3}. \label{eq:skew}\ee
Substitution of (\ref{eq:mean}-\ref{eq:var}) into (\ref{eq:skew}) provides a complicated expression for $\Sk$ which is in general nonzero.
Therefore we should generally expect the distribution of $u$ under proportional resetting to be non-Gaussian.

Evaluating weak and strong resetting limits in the skew formula (\ref{eq:skew}) can produce some insight into the behavior of the distribution.
The resulting formulas involve the combination
\be \Pe = \frac{\Gamma^2}{2 D r}.\ee
This is a P\'{e}clet number which measures the relative contribution of advection and diffusion to the evolution of $u$ between resets.
The average time between resets is $1/r$, so amplitude changes over this timescale due respectively to advection and diffusion are $U_A = \Gamma/r$ and $U_D = \pm \sqrt{2 D/r}$. The squared ratio of these quantities gives $\Pe = (U_A/U_D)^2$.

For strong resets, substituting the appropriate quantities into  (\ref{eq:skew}), expanding around $\ve \approx 0$, then expanding a second time for $\Pe^{-1} \approx 0$ produces
\be \Sk \approx 2 - \frac{3}{8}\Pe^{-2}.\ee
Therefore the skew coefficient approximates the characteristic value $\Sk = 2$ of a one-sided exponential distribution when resets remove the majority of $u$, at least provided $\Pe$ is large.

Similarly for weak resets, substitution, expansion around $\ve \approx 1 $, and a second expansion around $\Pe \approx 0$ produces
\be \Sk \approx \sqrt{\frac{2 \Pe}{1-\ve}}. \ee
Thus $\Sk$ is also capable of approaching the Gaussian extreme $\Sk \approx 0$ for resets which change $u$ very little, at least if $\Pe$ is small. Together these limits of the skew formula suggest that the distribution of $u$ can interpolate between Gaussian and exponential forms depending on the specific values of $\ve$ and $\Pe$. 

\subsection{Derivation of the steady-state distribution from the pantograph equation}

The Fokker-Planck equation (\ref{eq:fp}) is similar to functional equations originally devised to predict the deformation of overhead power delivery wires for electric trains (i.e., pantograph wires)  \cite{Fox1971,Ockendon1971}.
Analogues of (\ref{eq:fp}) have been used in biophysics to model the size distribution of a population of dividing cells \cite{Hall1989,VanBrunt2018,Efendiev2018}. Other contexts for pantograph-type functional equations are discussed in \cite{Zaidi2015,Zaidi2021}.

\begin{figure}
	\includegraphics{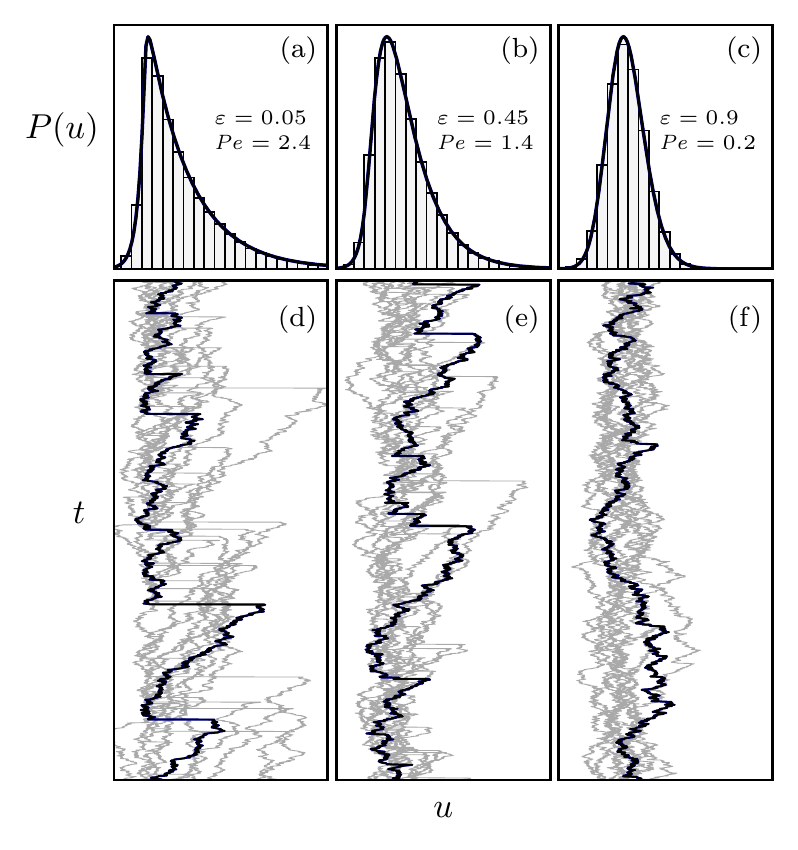}
	\caption{This figure depicts Monte Carlo simulations alongside the analytical distribution (\ref{eq:steadystate}) for conditions of strong (a,d), moderate (b,e), and weak (c,f) proportional resetting. A transition is evident in the distributions from a two-sided exponential-like shape for strong resetting, to a Gamma-like shape for moderate resetting, and eventually a Gaussian-like distribution for weak resetting. In all cases, (\ref{eq:steadystate}) shows excellent agreement with the numerical simulations. Simulations alternate the Euler-Mayarama method on $\dot{u}=\Gamma+\xi(t)$ with resets $u\rightarrow \ve u$, cf. \cite{Barik2006}. \label{fig:fig2}}
\end{figure}
The pantograph equation (\ref{eq:fp}) can be solved in steady-state conditions using the Fourier transform:
\be \tilde{P}(z) = \int_{\infty}^\infty e^{i u z} P(u) du. \label{eq:fourier}\ee
Introducing this transform to (\ref{eq:fp}) converts the functional differential equation to a recursion relation:
\be \tilde{P}(z) = \frac{\tilde{P}(z \ve)}{q(z)}. \label{eq:recurse} \ee
The polynomial $ q(z) = r^{-1}Dz^2 - i r^{-1}\Gamma z + 1$ is a quadratic function of the transform variable.
Recursing $N$ times provides
\be \tilde{P}(z) = \frac{\tilde{P}(z \ve^N)}{q(s \ve^0)q(s\ve^1)\dots q(s\ve^N)}.\label{eq:recursion}\ee
According to (\ref{eq:fourier}), the normalization condition for the distribution in Fourier space is $\tilde{P}(0)=1$. This implies $\lim_{N\rightarrow \infty} \tilde{P}(z \ve^N) = \tilde{P}(0)=1$. Taking the limit $N \rightarrow \infty$ in (\ref{eq:recursion}) and using this property gives 
\be \tilde{P}(z) =\frac{1}{q(z\ve^0)q(z\ve^1)q(z\ve^2) \dots} . \label{eq:formal}\ee
This transform can be inverted with some effort. The calculation is conducted in appendix \ref{sec:langsteadyderiv}, providing
\be
 P(u) = \sum_{\sigma=\pm}\sum_{l=0}^\infty \frac{ \Theta(-\sigma u) \ve^{-l}e^{\lambda_\sigma \ve^{-l}u}}{K_\sigma \prod_{m=1}^l q(i\lam_\sigma \ve^{-m}) } 
 \label{eq:steadystate}.
\ee
This result is a key contribution of the paper. It represents the steady-state distribution of the advection-diffusion process with proportional resetting. 

This solution is a type of Dirichlet series. These commonly arise as the eigenfunctions of pantograph equations \cite{Kim1998, VanBrunt2011}. 
In (\ref{eq:steadystate}), $\Theta(\cdot)$ is the Heaviside step function with the convention $\Theta(0)=1/2$. The $K_\pm$ are normalization factors. $i \lambda_\pm$ are the roots of the polynomial $q(z)$. Specific formulas for $\lam_\pm$ and $K_\pm$ are provided in Appendix \ref{sec:langsteadyderiv}.

The solution (\ref{eq:steadystate}) basically involves a sum of modes that decay from $u=0$ over scales that depend on the sign of $u$. The distribution can therefore exhibit asymmetry around $u=0$, and there is potential for a cusp singularity there, depending on the values of $\ve$, $\Gamma$, $D$, and $r$. The distribution always matches at $u=0$: $P(0^+)=P(0^-)$, and it is always normalized, as required.

The main behaviors of (\ref{eq:steadystate}) are illustrated in Figure \ref{fig:fig2}(a). Here it is also shown that the distribution (\ref{eq:steadystate}) shows excellent agreement with numerical simulations. 
The distribution appears visually similar to a two-sided exponential distribution when resets remove a large proportion of the amplitude, $\ve \rightarrow 0$. As resets become weaker, the distribution becomes Gamma- and eventually Gaussian-like, as shown in Figure \ref{fig:fig2}(b-c).
These observations are consistent with the skew calculations made before, although the limiting behaviors are also less sensitive to $\Pe$ than these calculations suggest.

\subsection{Exponential and Gaussian regimes}

\begin{figure}
	\includegraphics{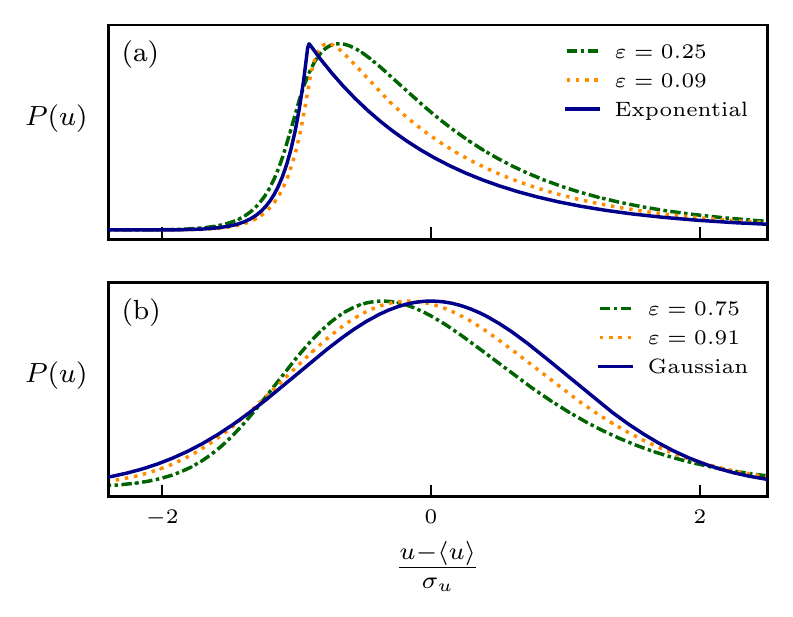}
	\caption{The distribution (\ref{eq:steadystate}) approaches two-sided exponential and Gaussian distributions for extreme values $\ve \rightarrow 0$ and $\ve \rightarrow 1$ of the reset proportion, representing respectively strong and weak resetting. Plotted distributions for moderate $\ve$ (green and orange) are computed using (\ref{eq:steadystate}). Plotted exponential and Gaussian distributions (blue) are formulas (\ref{eq:expon}) and (\ref{eq:gaussian}) respectively. In these calculations, $\Pe = 0.3$. The convergence is not sensitive to this value. \label{fig:fig3}}
\end{figure}

The exponential and Gaussian end-member behaviors of (\ref{eq:steadystate}) can be derived mathematically.
The exponential case for $\ve \rightarrow 0$ is particularly easy to see.
In this limit, the numerators of each term in the Dirichlet series scale as $\ve^{-l}  e^{-|\lambda_l u| \ve^{-l}}$, while the denominators scale as as $\ve^{-2 l}$.
When $\ve \rightarrow 0$, every term in the series excluding $l=0$ therefore vanishes at least as fast as $\ve^l$.
The distribution limits to
\be P(u) \approx \sum_{\sigma=\pm} \frac{\Theta(-\sigma u)}{K_\sigma}  \exp\big(-|\lambda_\sigma u| \big) + \mathcal{O}(\ve) \label{eq:expon}.\ee
This is a two-sided exponential distribution.
The approach toward this limit as $\ve$ vanishes is detailed in Fig. \ref{fig:fig3}(a).

The opposite limit of $\ve \rightarrow 1$ in (\ref{eq:steadystate}) is more difficult to evaluate. In this case, the denominator terms in the series all diverge.
To circumvent this issue it is useful to follow \cite{Hall1989} and introduce the scaled variable
\be 
z = \frac{ u - \bra u \ket }{\sigma_u}\label{eq:zscore} \ee
to the original Fokker-Planck equation (\ref{eq:fp}).
This involves the moment results (\ref{eq:mean}) and (\ref{eq:var}).
Since $\bra u \ket$ and $\sigma_u$ also diverge as $\ve \rightarrow 1$, the scaled distribution $Q(z) = \sigma_u P(u)$ turns out to be well-behaved in the limit.

As shown in Appendix \ref{sec:appb}, substituting (\ref{eq:zscore}) into (\ref{eq:fp}) gives a Fokker-Planck equation for $Q(z)$. Taking $\ve \rightarrow 1$ limits this to the well-known equation for the standard normal distribution, ultimately providing
\be P(u) \approx  \sqrt{\frac{1}{2\pi\sigma_u^2}}\exp\Bigg(-\frac{\big[u-\bra u \ket\big]^2}{2\sigma_u^2}\Bigg) + \mathcal{O}(1-\ve). \label{eq:gaussian}\ee
The approach to this limiting Gaussian distribution as resetting becomes weak is illustrated in Fig. \ref{fig:fig3}(b).

\section{Discussion}
\label{sec:disc}

\subsection{Next steps and relation to conventional resetting}

The Fokker-Planck equation (\ref{eq:fp}) introduces a modified resetting mechanism to an advection-diffusion process. This mechanism has a unique feature compared to the earlier resetting processes: the amount by which the underlying diffusion process is shifted during resets depends on its value.
This is a type of state-dependent resetting, similar to \cite{Roldan2017,Pinsky2020}, except the reset \textit{magnitude} depends on $u$, not the reset \textit{rate}.

It is straightforward to generalize (\ref{eq:fp}) to include more general state-dependent resetting magnitudes than the proportional resetting process developed here.
Considering that resets map $u$ to $u'$ through some function $u'= g(u)$, (\ref{eq:fp}) generalizes to
\begin{multline}
	\dot{P}(u,t) = \hat{L} P(u,t) -r P(u,t)\\ + r \int_{-\infty}^\infty du' P(u',t)\delta\big(u-g(u')\big). \label{eq:reset}
\end{multline}
It is only when $g(u) = \ve u$ that this simplifies to proportional resetting.
The original model of \cite{Evans2011,Evans2011a} is obtained when resets become state independent: $g(u) = u_r$.
The solution of resetting with other state-dependent magnitudes provides an interesting paradigm for future studies.

\subsection{First passage times of proportional resetting}

Given the significance of resetting to searching problems, an important next step will be to calculate the first passage characteristics of the proportional resetting process to a target.
One can formulate a backward Fokker-Planck equation using the same approach as before.
The backward equation can be linked to any first passage characteristics of interest \cite{Gardiner1983,Masoliver2018}.

Considering that the proportional resetting process starts at $u>0$ at $t=0$, and that it absorbs as soon as it hits $u=0$, the governing equation of the mean first passage time $T(u)$ to the origin turns out to be
\be -1 =  D T''(u) + \Gamma T'(u)  -r T(u) + r T(\ve u).\label{eq:mfpt}\ee
Here the appropriate boundary conditions are $T(0)=0$ and $T'(\infty)=0$: absorption at the origin and reflection at infinity.
Equation (\ref{eq:mfpt}) is functional, like (\ref{eq:fp}).  Unfortunately it is also far more challenging to solve than (\ref{eq:fp}) because it is inhomogeneous. Only for extreme resetting proportions ($\ve \rightarrow 0,1$) does solving this equation become straightforward.

For weak resets, the solution of (\ref{eq:mfpt}) diverges. This is to be expected as the problem limits to free Brownian motion \cite{Balakrishnan2021}.
For strong resets, (\ref{eq:mfpt}) can be expanded at $\ve \approx 0$ to remove its functional dependence.
The result solves for
\be T(u) = \frac{1-e^{-\lam_+ u}}{r}.\ee
When the process starts sufficiently far from the absorbing point, $u \gg 1/\lam_+$, the mean first passage time is the same as the timescale for a single reset: $T(u) \approx 1/r$.
When any reset takes $u$ to the origin, one reset is enough.

Future studies should evaluate the first passage characteristics of diffusion with other state-dependent reset magnitudes. The task is to generalize the pioneering works on optimal resetting \cite{Evans2011,Evans2013} to include different magnitudes than $g(u)=u_r$.
At this point we can only speculate that certain $g(u)$ may provide additional advantages for optimizing first passage characteristics.

\section{Conclusion}
\label{sec:conclusion}

This paper has detailed a stochastic process with resetting in which the reset magnitudes depend on the amplitude of the process.
This was formulated as a Langevin model with Gaussian white and Poisson shot noise terms.
The associated Fokker-Planck equation was derived as a functional differential equation, and the exact moments and steady-state distribution function were obtained.
The results show excellent agreement with numerical simulations.
The steady-state distribution adapts between exponential and Gaussian forms depending on the proportion of the amplitude lost during resets.

The proportional resetting model will be useful for applications in which a diffusive quantity is randomly reset to some fraction of its amount. Possible contexts include finance, physics, population modeling, and environmental science.
For processes which reset to some other function of their amount, (\ref{eq:reset}) is the required generalization.
Future works should study processes with more general state-dependent resetting magnitudes and evaluate their first passage properties.

\begin{acknowledgments}
Support is acknowledged from the University of British Columbia.
\end{acknowledgments}

\appendix
\section{Steady-state distribution}
\label{sec:langsteadyderiv}
To calculate the inverse Fourier transform of (\ref{eq:formal}) the overall strategy is to conduct a partial fractions expansion on the infinite product, then invert term-by-term.
The polynomial $q(z)$ can always be factored as 
\be q(z) = r^{-1}D(z - i\lambda_-)(z - i\lambda_+),\ee
with the factors
\be \lambda_\pm = \frac{\Gamma}{2D}\Big[1 \pm \sqrt{ 1 + 4D r/\Gamma^2} \Big]. \label{eq:lambdas}\ee
Using these factors to expand $\tilde{P}(z)$ in partial fractions provides
\be \tilde{P}(z)  =  \sum_{l=0}^\infty  \Big[ \frac{R_l^-}{ z\ve^l - i\lambda_-} + \frac{R_l^+}{z \ve^l - i\lambda_+} \Big] , \label{eq:expanded}\ee
where the coefficients $R_l^\pm$ are the residues (up to constant factors) of the product $[q(z\ve^0)q(z \ve^1)\dots]^{-1}$:
\be R_l^\pm = \frac{z \ve^{l} - i\lambda_\pm }{q(z \ve^0)q(z \ve^1)\dots } \Big|_{z= i \lambda_\pm \ve^{-l}} .\ee
Evaluating these coefficients and incorporating them in (\ref{eq:expanded}) produces 
\bwt
\begin{multline}
	 \tilde{P}(z) = \frac{-1}{i D(\lambda_+ - \lambda_-) \prod_{m=1}^\infty q(i \lambda_- \ve^m)} \sum_{l=0}^\infty \frac{1}{(z\ve^l-i\lambda_-)\prod_{m=1}^l q(i\lambda_- \ve^{-m})} 
	\\+ \frac{1}{i D(\lambda_+ - \lambda_-) \prod_{m=1}^\infty q(i \lambda_+ \ve^m)} \sum_{l=0}^\infty \frac{1}{(z\ve^l-i \lambda_+)\prod_{m=1}^l q(i\lambda_+ \ve^{-m})}.  \label{eq:invertme}
\end{multline}
\ewt
Empty products are considered unity.
Equation (\ref{eq:expanded}) can be inverted term by term with contour integration. The inverse Fourier transform is defined as $P(u) = \int_{-\infty}^\infty \frac{dz}{2\pi}e^{-i z u}\tilde{P}(z).$ Consider two closed square contours $C_+$ and $C_-$ that revolve counter-clockwise in the complex plane. Both contact on the $\Re$ axis. $C_+$ resides in quadrants I and II, while $C_-$ resides in III and IV. The contours have infinite extent. 

When inverting (\ref{eq:invertme}), the relevant contour for the resulting integrals depends on the sign of $u$. 
In terms of these contours, all integrals in the inversion have the form
 \be \int_{-\infty}^\infty \frac{d z}{2\pi i }\frac{e^{-i z u }}{z - i k} =  \Big[\Theta(-u)\oint_{C_+} -\Theta(u)\oint_{C_-}\Big] \frac{d\zeta}{2 \pi i} \frac{e^{-i \zeta u }}{\zeta - i k}.\ee
 The integrand has a first order pole at $\zeta = i k$.
 This pole is within either $C_+$ or $C_-$ depending on the sign of $k$.
 Evaluating the residues in each contour with attention to the sign of $k$ produces  
 \be \int_{-\infty}^\infty \frac{d z}{2\pi i }\frac{e^{-i z u }}{z - i k} = e^{- |u k| }\big[\Theta(k)\Theta(-u)-\Theta(-k)\Theta(u)\big]. \label{eq:contour}\ee
 Finally, inverting (\ref{eq:invertme}) with (\ref{eq:contour}) produces (\ref{eq:steadystate}).

 The normalization factors $K_\pm$ in (\ref{eq:steadystate}) are defined as
 \be K_\pm = r^{-1}D(\lam_+-\lam_-)\prod_{m=1}^\infty q(i\lam_\pm \ve^m).\ee
 Evaluating these infinite products is numerically impractical. 
 Instead, using the definition of the $q$-Pochhammer symbol \cite{Andrews1999}
\be (a;q)_n = \prod_{k=0}^n (1-a q^k) ,\ee
the normalization factors can be written
\be K_\pm = r^{-1} D (\lam_\pm- \lam_\mp) \frac{\big(\ve ;\ve\big)_\infty \big(\lam_\pm/\lam_\mp;\ve\big)_\infty}{1-\lam_\pm/\lam_\mp}. \label{eq:seriesrep}\ee
Using the series representation \cite{Andrews1999}
\be  (a;q)_\infty = \sum_{n=0}^\infty \frac{(-)^nq^{n(n-1)/2}}{(q;q)_n}a^n, \ee
formula (\ref{eq:seriesrep}) can then be approximated as a product of two series.

\section{Weak resetting}
\label{sec:appb}
Incorporating the definitions of $z$ and $Q(z)$ from the text into (\ref{eq:fp}) provides the scaled Fokker-Planck equation
\begin{multline}(1-\ve^2) \frac{D}{2D + \Gamma} 0 = Q''(z) - \frac{\Gamma\sqrt{1-\ve^2}}{\sqrt{2D+\Gamma^2}}Q'(z) - Q(z) \\+ \frac{1}{\ve} Q\Big(z + \Big[\frac{1-\ve}{\ve}z + \frac{\Gamma\sqrt{1-\ve^2}}{\ve\sqrt{2D+\Gamma^2}} \Big]\Big).\end{multline}
This equation remains exact and is only a change of variables from (\ref{eq:fp}).

Now all terms are expanded around $\ve \approx 1$ to obtain
\be 0 \approx Q(z) + z Q'(z) + Q''(z) + \mathcal{O}\big(\sqrt{1-\ve}\big).\ee
This is the classic Ornstein-Uhlenbeck Fokker-Planck equation whose solution is the standard normal distribution for $Q(z)$:
\be Q(z) \approx (2\pi)^{-1/2}\exp\big(-z^2\big) + \mathcal{O}\big(\sqrt{1-\ve}\big).\ee
This solution provides (\ref{eq:gaussian}) when transformed back to the original variables $P(u)$ and $u$.

\bibliography{biblio.bib}

\begin{thebibliography}{58}%
\makeatletter
\providecommand \@ifxundefined [1]{%
 \@ifx{#1\undefined}
}%
\providecommand \@ifnum [1]{%
 \ifnum #1\expandafter \@firstoftwo
 \else \expandafter \@secondoftwo
 \fi
}%
\providecommand \@ifx [1]{%
 \ifx #1\expandafter \@firstoftwo
 \else \expandafter \@secondoftwo
 \fi
}%
\providecommand \natexlab [1]{#1}%
\providecommand \enquote  [1]{``#1''}%
\providecommand \bibnamefont  [1]{#1}%
\providecommand \bibfnamefont [1]{#1}%
\providecommand \citenamefont [1]{#1}%
\providecommand \href@noop [0]{\@secondoftwo}%
\providecommand \href [0]{\begingroup \@sanitize@url \@href}%
\providecommand \@href[1]{\@@startlink{#1}\@@href}%
\providecommand \@@href[1]{\endgroup#1\@@endlink}%
\providecommand \@sanitize@url [0]{\catcode `\\12\catcode `\$12\catcode
  `\&12\catcode `\#12\catcode `\^12\catcode `\_12\catcode `\%12\relax}%
\providecommand \@@startlink[1]{}%
\providecommand \@@endlink[0]{}%
\providecommand \url  [0]{\begingroup\@sanitize@url \@url }%
\providecommand \@url [1]{\endgroup\@href {#1}{\urlprefix }}%
\providecommand \urlprefix  [0]{URL }%
\providecommand \Eprint [0]{\href }%
\providecommand \doibase [0]{https://doi.org/}%
\providecommand \selectlanguage [0]{\@gobble}%
\providecommand \bibinfo  [0]{\@secondoftwo}%
\providecommand \bibfield  [0]{\@secondoftwo}%
\providecommand \translation [1]{[#1]}%
\providecommand \BibitemOpen [0]{}%
\providecommand \bibitemStop [0]{}%
\providecommand \bibitemNoStop [0]{.\EOS\space}%
\providecommand \EOS [0]{\spacefactor3000\relax}%
\providecommand \BibitemShut  [1]{\csname bibitem#1\endcsname}%
\let\auto@bib@innerbib\@empty
\bibitem [{\citenamefont {Montero}\ \emph {et~al.}(2017)\citenamefont
  {Montero}, \citenamefont {Mas{\'{o}}-Puigdellosas},\ and\ \citenamefont
  {Villarroel}}]{Montero2017}%
  \BibitemOpen
  \bibfield  {author} {\bibinfo {author} {\bibfnamefont {M.}~\bibnamefont
  {Montero}}, \bibinfo {author} {\bibfnamefont {A.}~\bibnamefont
  {Mas{\'{o}}-Puigdellosas}},\ and\ \bibinfo {author} {\bibfnamefont
  {J.}~\bibnamefont {Villarroel}},\ }\bibfield  {title} {\bibinfo {title}
  {{Continuous-time random walks with reset events: Historical background and
  new perspectives}},\ }\href {https://doi.org/10.1140/epjb/e2017-80348-4}
  {\bibfield  {journal} {\bibinfo  {journal} {European Physical Journal B}\
  }\textbf {\bibinfo {volume} {90}},\ \bibinfo {pages} {1} (\bibinfo {year}
  {2017})},\ \Eprint {https://arxiv.org/abs/1706.04812} {arXiv:1706.04812}
  \BibitemShut {NoStop}%
\bibitem [{\citenamefont {Evans}\ \emph {et~al.}(2020)\citenamefont {Evans},
  \citenamefont {Majumdar},\ and\ \citenamefont {Schehr}}]{Evans2020}%
  \BibitemOpen
  \bibfield  {author} {\bibinfo {author} {\bibfnamefont {M.~R.}\ \bibnamefont
  {Evans}}, \bibinfo {author} {\bibfnamefont {S.~N.}\ \bibnamefont
  {Majumdar}},\ and\ \bibinfo {author} {\bibfnamefont {G.}~\bibnamefont
  {Schehr}},\ }\bibfield  {title} {\bibinfo {title} {{Stochastic resetting and
  applications}},\ }\bibfield  {journal} {\bibinfo  {journal} {Journal of
  Physics A: Mathematical and Theoretical}\ }\textbf {\bibinfo {volume} {53}},\
  \href {https://doi.org/10.1088/1751-8121/ab7cfe} {10.1088/1751-8121/ab7cfe}
  (\bibinfo {year} {2020}),\ \Eprint {https://arxiv.org/abs/1910.07993}
  {arXiv:1910.07993} \BibitemShut {NoStop}%
\bibitem [{\citenamefont {Eule}\ and\ \citenamefont
  {Metzger}(2016)}]{Eule2016}%
  \BibitemOpen
  \bibfield  {author} {\bibinfo {author} {\bibfnamefont {S.}~\bibnamefont
  {Eule}}\ and\ \bibinfo {author} {\bibfnamefont {J.~J.}\ \bibnamefont
  {Metzger}},\ }\bibfield  {title} {\bibinfo {title} {{Non-equilibrium steady
  states of stochastic processes with intermittent resetting}},\ }\bibfield
  {journal} {\bibinfo  {journal} {New Journal of Physics}\ }\textbf {\bibinfo
  {volume} {18}},\ \href {https://doi.org/10.1088/1367-2630/18/3/033006}
  {10.1088/1367-2630/18/3/033006} (\bibinfo {year} {2016})\BibitemShut
  {NoStop}%
\bibitem [{\citenamefont {Evans}\ and\ \citenamefont
  {Majumdar}(2011{\natexlab{a}})}]{Evans2011a}%
  \BibitemOpen
  \bibfield  {author} {\bibinfo {author} {\bibfnamefont {M.~R.}\ \bibnamefont
  {Evans}}\ and\ \bibinfo {author} {\bibfnamefont {S.~N.}\ \bibnamefont
  {Majumdar}},\ }\bibfield  {title} {\bibinfo {title} {{Diffusion with
  stochastic resetting}},\ }\href
  {https://doi.org/10.1103/PhysRevLett.106.160601} {\bibfield  {journal}
  {\bibinfo  {journal} {Physical Review Letters}\ }\textbf {\bibinfo {volume}
  {106}},\ \bibinfo {pages} {1} (\bibinfo {year} {2011}{\natexlab{a}})},\
  \Eprint {https://arxiv.org/abs/1102.2704} {arXiv:1102.2704} \BibitemShut
  {NoStop}%
\bibitem [{\citenamefont {Evans}\ \emph {et~al.}(2013)\citenamefont {Evans},
  \citenamefont {Majumdar},\ and\ \citenamefont {Mallick}}]{Evans2013}%
  \BibitemOpen
  \bibfield  {author} {\bibinfo {author} {\bibfnamefont {M.~R.}\ \bibnamefont
  {Evans}}, \bibinfo {author} {\bibfnamefont {S.~N.}\ \bibnamefont
  {Majumdar}},\ and\ \bibinfo {author} {\bibfnamefont {K.}~\bibnamefont
  {Mallick}},\ }\bibfield  {title} {\bibinfo {title} {{Optimal diffusive
  search: Nonequilibrium resetting versus equilibrium dynamics}},\ }\bibfield
  {journal} {\bibinfo  {journal} {Journal of Physics A: Mathematical and
  Theoretical}\ }\textbf {\bibinfo {volume} {46}},\ \href
  {https://doi.org/10.1088/1751-8113/46/18/185001}
  {10.1088/1751-8113/46/18/185001} (\bibinfo {year} {2013}),\ \Eprint
  {https://arxiv.org/abs/1212.4096} {arXiv:1212.4096} \BibitemShut {NoStop}%
\bibitem [{\citenamefont {Evans}\ and\ \citenamefont
  {Majumdar}(2011{\natexlab{b}})}]{Evans2011}%
  \BibitemOpen
  \bibfield  {author} {\bibinfo {author} {\bibfnamefont {M.~R.}\ \bibnamefont
  {Evans}}\ and\ \bibinfo {author} {\bibfnamefont {S.~N.}\ \bibnamefont
  {Majumdar}},\ }\bibfield  {title} {\bibinfo {title} {{Diffusion with optimal
  resetting}},\ }\href {https://doi.org/10.1088/1751-8113/44/43/435001}
  {\bibfield  {journal} {\bibinfo  {journal} {Journal of Physics A:
  Mathematical and Theoretical}\ }\textbf {\bibinfo {volume} {44}},\ \bibinfo
  {pages} {1} (\bibinfo {year} {2011}{\natexlab{b}})},\ \Eprint
  {https://arxiv.org/abs/1107.4225} {arXiv:1107.4225} \BibitemShut {NoStop}%
\bibitem [{\citenamefont {Majumdar}\ \emph {et~al.}(2015)\citenamefont
  {Majumdar}, \citenamefont {Sabhapandit},\ and\ \citenamefont
  {Schehr}}]{Majumdar2015}%
  \BibitemOpen
  \bibfield  {author} {\bibinfo {author} {\bibfnamefont {S.~N.}\ \bibnamefont
  {Majumdar}}, \bibinfo {author} {\bibfnamefont {S.}~\bibnamefont
  {Sabhapandit}},\ and\ \bibinfo {author} {\bibfnamefont {G.}~\bibnamefont
  {Schehr}},\ }\bibfield  {title} {\bibinfo {title} {{Random walk with random
  resetting to the maximum position}},\ }\href
  {https://doi.org/10.1103/PhysRevE.92.052126} {\bibfield  {journal} {\bibinfo
  {journal} {Physical Review E - Statistical, Nonlinear, and Soft Matter
  Physics}\ }\textbf {\bibinfo {volume} {92}},\ \bibinfo {pages} {1} (\bibinfo
  {year} {2015})},\ \Eprint {https://arxiv.org/abs/1509.04516}
  {arXiv:1509.04516} \BibitemShut {NoStop}%
\bibitem [{\citenamefont {Boyer}\ and\ \citenamefont
  {Solis-Salas}(2014)}]{Boyer2014}%
  \BibitemOpen
  \bibfield  {author} {\bibinfo {author} {\bibfnamefont {D.}~\bibnamefont
  {Boyer}}\ and\ \bibinfo {author} {\bibfnamefont {C.}~\bibnamefont
  {Solis-Salas}},\ }\bibfield  {title} {\bibinfo {title} {{Random walks with
  preferential relocations to places visited in the past and their application
  to biology}},\ }\href {https://doi.org/10.1103/PhysRevLett.112.240601}
  {\bibfield  {journal} {\bibinfo  {journal} {Physical Review Letters}\
  }\textbf {\bibinfo {volume} {112}},\ \bibinfo {pages} {1} (\bibinfo {year}
  {2014})},\ \Eprint {https://arxiv.org/abs/1403.6069} {arXiv:1403.6069}
  \BibitemShut {NoStop}%
\bibitem [{\citenamefont {Rold{\'{a}}n}\ and\ \citenamefont
  {Gupta}(2017)}]{Roldan2017}%
  \BibitemOpen
  \bibfield  {author} {\bibinfo {author} {\bibfnamefont {{\'{E}}.}~\bibnamefont
  {Rold{\'{a}}n}}\ and\ \bibinfo {author} {\bibfnamefont {S.}~\bibnamefont
  {Gupta}},\ }\bibfield  {title} {\bibinfo {title} {{Path-integral formalism
  for stochastic resetting: Exactly solved examples and shortcuts to
  confinement}},\ }\bibfield  {journal} {\bibinfo  {journal} {Physical Review
  E}\ }\textbf {\bibinfo {volume} {96}},\ \href
  {https://doi.org/10.1103/PhysRevE.96.022130} {10.1103/PhysRevE.96.022130}
  (\bibinfo {year} {2017}),\ \Eprint {https://arxiv.org/abs/1703.10615}
  {arXiv:1703.10615} \BibitemShut {NoStop}%
\bibitem [{\citenamefont {Pinsky}(2020)}]{Pinsky2020}%
  \BibitemOpen
  \bibfield  {author} {\bibinfo {author} {\bibfnamefont {R.~G.}\ \bibnamefont
  {Pinsky}},\ }\bibfield  {title} {\bibinfo {title} {{Diffusive search with
  spatially dependent resetting}},\ }\href
  {https://doi.org/10.1016/j.spa.2019.08.008} {\bibfield  {journal} {\bibinfo
  {journal} {Stochastic Processes and their Applications}\ }\textbf {\bibinfo
  {volume} {130}},\ \bibinfo {pages} {2954} (\bibinfo {year} {2020})},\ \Eprint
  {https://arxiv.org/abs/1805.00320} {arXiv:1805.00320} \BibitemShut {NoStop}%
\bibitem [{\citenamefont {Pal}\ \emph {et~al.}(2016)\citenamefont {Pal},
  \citenamefont {Kundu},\ and\ \citenamefont {Evans}}]{Pal2016}%
  \BibitemOpen
  \bibfield  {author} {\bibinfo {author} {\bibfnamefont {A.}~\bibnamefont
  {Pal}}, \bibinfo {author} {\bibfnamefont {A.}~\bibnamefont {Kundu}},\ and\
  \bibinfo {author} {\bibfnamefont {M.~R.}\ \bibnamefont {Evans}},\ }\bibfield
  {title} {\bibinfo {title} {{Diffusion under time-dependent resetting}},\
  }\href {https://doi.org/10.1088/1751-8113/49/22/225001} {\bibfield  {journal}
  {\bibinfo  {journal} {Journal of Physics A: Mathematical and Theoretical}\
  }\textbf {\bibinfo {volume} {49}},\ \bibinfo {pages} {1} (\bibinfo {year}
  {2016})},\ \Eprint {https://arxiv.org/abs/1512.08211} {arXiv:1512.08211}
  \BibitemShut {NoStop}%
\bibitem [{\citenamefont {Shkilev}(2017)}]{Shkilev2017}%
  \BibitemOpen
  \bibfield  {author} {\bibinfo {author} {\bibfnamefont {V.~P.}\ \bibnamefont
  {Shkilev}},\ }\bibfield  {title} {\bibinfo {title} {{Continuous-time random
  walk under time-dependent resetting}},\ }\href
  {https://doi.org/10.1103/PhysRevE.96.012126} {\bibfield  {journal} {\bibinfo
  {journal} {Physical Review E}\ }\textbf {\bibinfo {volume} {96}},\ \bibinfo
  {pages} {1} (\bibinfo {year} {2017})}\BibitemShut {NoStop}%
\bibitem [{\citenamefont {Mas{\'{o}}-Puigdellosas}\ \emph
  {et~al.}(2019)\citenamefont {Mas{\'{o}}-Puigdellosas}, \citenamefont
  {Campos},\ and\ \citenamefont {M{\'{e}}ndez}}]{Maso-Puigdellosas2019}%
  \BibitemOpen
  \bibfield  {author} {\bibinfo {author} {\bibfnamefont {A.}~\bibnamefont
  {Mas{\'{o}}-Puigdellosas}}, \bibinfo {author} {\bibfnamefont
  {D.}~\bibnamefont {Campos}},\ and\ \bibinfo {author} {\bibfnamefont
  {V.}~\bibnamefont {M{\'{e}}ndez}},\ }\bibfield  {title} {\bibinfo {title}
  {{Stochastic movement subject to a reset-and-residence mechanism: Transport
  properties and first arrival statistics}},\ }\href
  {https://doi.org/10.1088/1742-5468/ab02f3} {\bibfield  {journal} {\bibinfo
  {journal} {Journal of Statistical Mechanics: Theory and Experiment}\ }\textbf
  {\bibinfo {volume} {2019}},\ \bibinfo {pages} {1} (\bibinfo {year} {2019})},\
  \Eprint {https://arxiv.org/abs/1811.06742} {arXiv:1811.06742} \BibitemShut
  {NoStop}%
\bibitem [{\citenamefont {Bodrova}\ and\ \citenamefont
  {Sokolov}(2020)}]{Bodrova2020}%
  \BibitemOpen
  \bibfield  {author} {\bibinfo {author} {\bibfnamefont {A.~S.}\ \bibnamefont
  {Bodrova}}\ and\ \bibinfo {author} {\bibfnamefont {I.~M.}\ \bibnamefont
  {Sokolov}},\ }\bibfield  {title} {\bibinfo {title} {{Resetting processes with
  noninstantaneous return}},\ }\href
  {https://doi.org/10.1103/PhysRevE.101.052130} {\bibfield  {journal} {\bibinfo
   {journal} {Physical Review E}\ }\textbf {\bibinfo {volume} {101}},\ \bibinfo
  {pages} {1} (\bibinfo {year} {2020})},\ \Eprint
  {https://arxiv.org/abs/1907.12326} {arXiv:1907.12326} \BibitemShut {NoStop}%
\bibitem [{\citenamefont {Pal}\ \emph {et~al.}(2019)\citenamefont {Pal},
  \citenamefont {Ku{\'{s}}mierz},\ and\ \citenamefont {Reuveni}}]{Pal2019}%
  \BibitemOpen
  \bibfield  {author} {\bibinfo {author} {\bibfnamefont {A.}~\bibnamefont
  {Pal}}, \bibinfo {author} {\bibfnamefont {{\L}.}~\bibnamefont
  {Ku{\'{s}}mierz}},\ and\ \bibinfo {author} {\bibfnamefont {S.}~\bibnamefont
  {Reuveni}},\ }\bibfield  {title} {\bibinfo {title} {{Time-dependent density
  of diffusion with stochastic resetting is invariant to return speed}},\
  }\href {https://doi.org/10.1103/PhysRevE.100.040101} {\bibfield  {journal}
  {\bibinfo  {journal} {Physical Review E}\ }\textbf {\bibinfo {volume}
  {100}},\ \bibinfo {pages} {1} (\bibinfo {year} {2019})},\ \Eprint
  {https://arxiv.org/abs/arXiv:1907.12208v1} {arXiv:arXiv:1907.12208v1}
  \BibitemShut {NoStop}%
\bibitem [{\citenamefont {Pal}(2015)}]{Pal2015}%
  \BibitemOpen
  \bibfield  {author} {\bibinfo {author} {\bibfnamefont {A.}~\bibnamefont
  {Pal}},\ }\bibfield  {title} {\bibinfo {title} {{Diffusion in a potential
  landscape with stochastic resetting}},\ }\href
  {https://doi.org/10.1103/PhysRevE.91.012113} {\bibfield  {journal} {\bibinfo
  {journal} {Physical Review E - Statistical, Nonlinear, and Soft Matter
  Physics}\ }\textbf {\bibinfo {volume} {91}},\ \bibinfo {pages} {1} (\bibinfo
  {year} {2015})},\ \Eprint {https://arxiv.org/abs/1408.1230} {arXiv:1408.1230}
  \BibitemShut {NoStop}%
\bibitem [{\citenamefont {Chatterjee}\ \emph {et~al.}(2018)\citenamefont
  {Chatterjee}, \citenamefont {Christou},\ and\ \citenamefont
  {Schadschneider}}]{Chatterjee2018}%
  \BibitemOpen
  \bibfield  {author} {\bibinfo {author} {\bibfnamefont {A.}~\bibnamefont
  {Chatterjee}}, \bibinfo {author} {\bibfnamefont {C.}~\bibnamefont
  {Christou}},\ and\ \bibinfo {author} {\bibfnamefont {A.}~\bibnamefont
  {Schadschneider}},\ }\bibfield  {title} {\bibinfo {title} {{Diffusion with
  resetting inside a circle}},\ }\href
  {https://doi.org/10.1103/PhysRevE.97.062106} {\bibfield  {journal} {\bibinfo
  {journal} {Physical Review E}\ }\textbf {\bibinfo {volume} {97}},\ \bibinfo
  {pages} {1} (\bibinfo {year} {2018})},\ \Eprint
  {https://arxiv.org/abs/1801.09971} {arXiv:1801.09971} \BibitemShut {NoStop}%
\bibitem [{\citenamefont {Durang}\ \emph {et~al.}(2020)\citenamefont {Durang},
  \citenamefont {Lee}, \citenamefont {Lizana},\ and\ \citenamefont
  {Jeon}}]{Durang2020}%
  \BibitemOpen
  \bibfield  {author} {\bibinfo {author} {\bibfnamefont {X.}~\bibnamefont
  {Durang}}, \bibinfo {author} {\bibfnamefont {S.}~\bibnamefont {Lee}},
  \bibinfo {author} {\bibfnamefont {L.}~\bibnamefont {Lizana}},\ and\ \bibinfo
  {author} {\bibfnamefont {J.-H.}\ \bibnamefont {Jeon}},\ }\bibfield  {title}
  {\bibinfo {title} {{First-passage statistics under stochastic resetting in
  bounded domains}},\ }\href {https://doi.org/10.1088/1751-8121/ab15f5}
  {\bibfield  {journal} {\bibinfo  {journal} {Journal of Physics A:
  Mathematical and Theoretical}\ }\textbf {\bibinfo {volume} {52}},\ \bibinfo
  {pages} {1} (\bibinfo {year} {2020})}\BibitemShut {NoStop}%
\bibitem [{\citenamefont {Nagar}\ and\ \citenamefont
  {Gupta}(2016)}]{Nagar2016}%
  \BibitemOpen
  \bibfield  {author} {\bibinfo {author} {\bibfnamefont {A.}~\bibnamefont
  {Nagar}}\ and\ \bibinfo {author} {\bibfnamefont {S.}~\bibnamefont {Gupta}},\
  }\bibfield  {title} {\bibinfo {title} {{Diffusion with stochastic resetting
  at power-law times}},\ }\href {https://doi.org/10.1103/PhysRevE.93.060102}
  {\bibfield  {journal} {\bibinfo  {journal} {Physical Review E}\ }\textbf
  {\bibinfo {volume} {93}},\ \bibinfo {pages} {1} (\bibinfo {year} {2016})},\
  \Eprint {https://arxiv.org/abs/1512.02092} {arXiv:1512.02092} \BibitemShut
  {NoStop}%
\bibitem [{\citenamefont {Gupta}\ and\ \citenamefont
  {Nagar}(2016)}]{Gupta2016}%
  \BibitemOpen
  \bibfield  {author} {\bibinfo {author} {\bibfnamefont {S.}~\bibnamefont
  {Gupta}}\ and\ \bibinfo {author} {\bibfnamefont {A.}~\bibnamefont {Nagar}},\
  }\bibfield  {title} {\bibinfo {title} {{Resetting of fluctuating interfaces
  at power-law times}},\ }\href
  {https://doi.org/10.1088/1751-8113/49/44/445001} {\bibfield  {journal}
  {\bibinfo  {journal} {Journal of Physics A: Mathematical and Theoretical}\
  }\textbf {\bibinfo {volume} {49}},\ \bibinfo {pages} {1} (\bibinfo {year}
  {2016})},\ \Eprint {https://arxiv.org/abs/1604.06627} {arXiv:1604.06627}
  \BibitemShut {NoStop}%
\bibitem [{\citenamefont {Durang}\ \emph {et~al.}(2014)\citenamefont {Durang},
  \citenamefont {Henkel},\ and\ \citenamefont {Park}}]{Durang2014}%
  \BibitemOpen
  \bibfield  {author} {\bibinfo {author} {\bibfnamefont {X.}~\bibnamefont
  {Durang}}, \bibinfo {author} {\bibfnamefont {M.}~\bibnamefont {Henkel}},\
  and\ \bibinfo {author} {\bibfnamefont {H.}~\bibnamefont {Park}},\ }\bibfield
  {title} {\bibinfo {title} {{The statistical mechanics of the
  coagulation-diffusion process with a stochastic reset}},\ }\bibfield
  {journal} {\bibinfo  {journal} {Journal of Physics A: Mathematical and
  Theoretical}\ }\textbf {\bibinfo {volume} {47}},\ \href
  {https://doi.org/10.1088/1751-8113/47/4/045002}
  {10.1088/1751-8113/47/4/045002} (\bibinfo {year} {2014})\BibitemShut
  {NoStop}%
\bibitem [{\citenamefont {Sousa}\ and\ \citenamefont {Das}(2018)}]{Sousa2018}%
  \BibitemOpen
  \bibfield  {author} {\bibinfo {author} {\bibfnamefont {E.}~\bibnamefont
  {Sousa}}\ and\ \bibinfo {author} {\bibfnamefont {A.~K.}\ \bibnamefont
  {Das}},\ }\bibfield  {title} {\bibinfo {title} {{A fractional diffusion
  process with resetting}},\ }in\ \href
  {https://doi.org/10.1007/978-3-030-11539-5_59} {\emph {\bibinfo {booktitle}
  {Finite Difference Methods: Theory and Applications}}}\ (\bibinfo {address}
  {Lozenetz, Bulgaria},\ \bibinfo {year} {2018})\ pp.\ \bibinfo {pages}
  {509--516}\BibitemShut {NoStop}%
\bibitem [{\citenamefont {Evans}\ and\ \citenamefont
  {Majumdar}(2018)}]{Evans2018}%
  \BibitemOpen
  \bibfield  {author} {\bibinfo {author} {\bibfnamefont {M.~R.}\ \bibnamefont
  {Evans}}\ and\ \bibinfo {author} {\bibfnamefont {S.~N.}\ \bibnamefont
  {Majumdar}},\ }\bibfield  {title} {\bibinfo {title} {{Run and tumble particle
  under resetting: A renewal approach}},\ }\bibfield  {journal} {\bibinfo
  {journal} {Journal of Physics A: Mathematical and Theoretical}\ }\textbf
  {\bibinfo {volume} {51}},\ \href {https://doi.org/10.1088/1751-8121/aae74e}
  {10.1088/1751-8121/aae74e} (\bibinfo {year} {2018}),\ \Eprint
  {https://arxiv.org/abs/1808.06450} {arXiv:1808.06450} \BibitemShut {NoStop}%
\bibitem [{\citenamefont {Masoliver}(2019)}]{Masoliver2019}%
  \BibitemOpen
  \bibfield  {author} {\bibinfo {author} {\bibfnamefont {J.}~\bibnamefont
  {Masoliver}},\ }\bibfield  {title} {\bibinfo {title} {{Telegraphic processes
  with stochastic resetting}},\ }\href
  {https://doi.org/10.1103/PhysRevE.99.012121} {\bibfield  {journal} {\bibinfo
  {journal} {Physical Review E}\ }\textbf {\bibinfo {volume} {99}},\ \bibinfo
  {pages} {12121} (\bibinfo {year} {2019})}\BibitemShut {NoStop}%
\bibitem [{\citenamefont {Basu}\ \emph {et~al.}(2019)\citenamefont {Basu},
  \citenamefont {Kundu},\ and\ \citenamefont {Pal}}]{Basu2019}%
  \BibitemOpen
  \bibfield  {author} {\bibinfo {author} {\bibfnamefont {U.}~\bibnamefont
  {Basu}}, \bibinfo {author} {\bibfnamefont {A.}~\bibnamefont {Kundu}},\ and\
  \bibinfo {author} {\bibfnamefont {A.}~\bibnamefont {Pal}},\ }\bibfield
  {title} {\bibinfo {title} {{Symmetric exclusion process under stochastic
  resetting}},\ }\href {https://doi.org/10.1103/PhysRevE.100.032136} {\bibfield
   {journal} {\bibinfo  {journal} {Physical Review E}\ }\textbf {\bibinfo
  {volume} {100}},\ \bibinfo {pages} {1} (\bibinfo {year} {2019})},\ \Eprint
  {https://arxiv.org/abs/1906.11801} {arXiv:1906.11801} \BibitemShut {NoStop}%
\bibitem [{\citenamefont {Magoni}\ \emph {et~al.}(2020)\citenamefont {Magoni},
  \citenamefont {Majumdar},\ and\ \citenamefont {Schehr}}]{Magoni2020}%
  \BibitemOpen
  \bibfield  {author} {\bibinfo {author} {\bibfnamefont {M.}~\bibnamefont
  {Magoni}}, \bibinfo {author} {\bibfnamefont {S.~N.}\ \bibnamefont
  {Majumdar}},\ and\ \bibinfo {author} {\bibfnamefont {G.}~\bibnamefont
  {Schehr}},\ }\bibfield  {title} {\bibinfo {title} {{Ising model with
  stochastic resetting}},\ }\bibfield  {journal} {\bibinfo  {journal} {Physical
  Review Research}\ }\textbf {\bibinfo {volume} {2}},\ \href
  {https://doi.org/10.1103/PhysRevResearch.2.033182}
  {10.1103/PhysRevResearch.2.033182} (\bibinfo {year} {2020}),\ \Eprint
  {https://arxiv.org/abs/2002.04867} {arXiv:2002.04867} \BibitemShut {NoStop}%
\bibitem [{\citenamefont {Brilliantov}\ and\ \citenamefont
  {Poschel}(2004)}]{Brilliantov2004}%
  \BibitemOpen
  \bibfield  {author} {\bibinfo {author} {\bibfnamefont {N.~V.}\ \bibnamefont
  {Brilliantov}}\ and\ \bibinfo {author} {\bibfnamefont {T.}~\bibnamefont
  {Poschel}},\ }\href
  {https://doi.org/10.1093/acprof:oso/9780198530381.001.0001} {\emph {\bibinfo
  {title} {{Kinetic Theory of Granular Gases}}}},\ \bibinfo {edition} {1st}\
  ed.\ (\bibinfo  {publisher} {Oxford University Press},\ \bibinfo {address}
  {Oxford},\ \bibinfo {year} {2004})\BibitemShut {NoStop}%
\bibitem [{\citenamefont {Suweis}\ \emph {et~al.}(2011)\citenamefont {Suweis},
  \citenamefont {Porporato}, \citenamefont {Rinaldo},\ and\ \citenamefont
  {Maritan}}]{Suweis2011}%
  \BibitemOpen
  \bibfield  {author} {\bibinfo {author} {\bibfnamefont {S.}~\bibnamefont
  {Suweis}}, \bibinfo {author} {\bibfnamefont {A.}~\bibnamefont {Porporato}},
  \bibinfo {author} {\bibfnamefont {A.}~\bibnamefont {Rinaldo}},\ and\ \bibinfo
  {author} {\bibfnamefont {A.}~\bibnamefont {Maritan}},\ }\bibfield  {title}
  {\bibinfo {title} {{Prescription-induced jump distributions in multiplicative
  Poisson processes}},\ }\href {https://doi.org/10.1103/PhysRevE.83.061119}
  {\bibfield  {journal} {\bibinfo  {journal} {Physical Review E}\ }\textbf
  {\bibinfo {volume} {83}},\ \bibinfo {pages} {1} (\bibinfo {year}
  {2011})}\BibitemShut {NoStop}%
\bibitem [{\citenamefont {Suweis}\ \emph {et~al.}(2010)\citenamefont {Suweis},
  \citenamefont {Rinaldo}, \citenamefont {{Van Der Zee}}, \citenamefont {Daly},
  \citenamefont {Maritan},\ and\ \citenamefont {Porporato}}]{Suweis2010}%
  \BibitemOpen
  \bibfield  {author} {\bibinfo {author} {\bibfnamefont {S.}~\bibnamefont
  {Suweis}}, \bibinfo {author} {\bibfnamefont {A.}~\bibnamefont {Rinaldo}},
  \bibinfo {author} {\bibfnamefont {S.~E.}\ \bibnamefont {{Van Der Zee}}},
  \bibinfo {author} {\bibfnamefont {E.}~\bibnamefont {Daly}}, \bibinfo {author}
  {\bibfnamefont {A.}~\bibnamefont {Maritan}},\ and\ \bibinfo {author}
  {\bibfnamefont {A.}~\bibnamefont {Porporato}},\ }\bibfield  {title} {\bibinfo
  {title} {{Stochastic modeling of soil salinity}},\ }\href
  {https://doi.org/10.1029/2010GL042495} {\bibfield  {journal} {\bibinfo
  {journal} {Geophysical Research Letters}\ }\textbf {\bibinfo {volume} {37}},\
  \bibinfo {pages} {1} (\bibinfo {year} {2010})}\BibitemShut {NoStop}%
\bibitem [{\citenamefont {Mau}\ \emph {et~al.}(2014)\citenamefont {Mau},
  \citenamefont {Feng},\ and\ \citenamefont {Porporato}}]{Mau2014}%
  \BibitemOpen
  \bibfield  {author} {\bibinfo {author} {\bibfnamefont {Y.}~\bibnamefont
  {Mau}}, \bibinfo {author} {\bibfnamefont {X.}~\bibnamefont {Feng}},\ and\
  \bibinfo {author} {\bibfnamefont {A.}~\bibnamefont {Porporato}},\ }\bibfield
  {title} {\bibinfo {title} {{Multiplicative jump processes and applications to
  leaching of salt and contaminants in the soil}},\ }\href
  {https://doi.org/10.1103/PhysRevE.90.052128} {\bibfield  {journal} {\bibinfo
  {journal} {Physical Review E - Statistical, Nonlinear, and Soft Matter
  Physics}\ }\textbf {\bibinfo {volume} {90}},\ \bibinfo {pages} {1} (\bibinfo
  {year} {2014})}\BibitemShut {NoStop}%
\bibitem [{\citenamefont {Clark}(1989)}]{Clark1989}%
  \BibitemOpen
  \bibfield  {author} {\bibinfo {author} {\bibfnamefont {J.~S.}\ \bibnamefont
  {Clark}},\ }\bibfield  {title} {\bibinfo {title} {{Ecological Disturbance as
  a Renewal Process: Theory and Application to Fire History}},\ }\href
  {https://doi.org/10.2307/3566083} {\bibfield  {journal} {\bibinfo  {journal}
  {Oikos}\ }\textbf {\bibinfo {volume} {56}},\ \bibinfo {pages} {17} (\bibinfo
  {year} {1989})}\BibitemShut {NoStop}%
\bibitem [{\citenamefont {Odorico}\ \emph {et~al.}(2006)\citenamefont
  {Odorico}, \citenamefont {Laio}, \citenamefont {Ridolfi}, \citenamefont
  {Odorico}, \citenamefont {Laio},\ and\ \citenamefont
  {Ridolfi}}]{Odorico2006a}%
  \BibitemOpen
  \bibfield  {author} {\bibinfo {author} {\bibfnamefont {P.~D.}\ \bibnamefont
  {Odorico}}, \bibinfo {author} {\bibfnamefont {F.}~\bibnamefont {Laio}},
  \bibinfo {author} {\bibfnamefont {L.}~\bibnamefont {Ridolfi}}, \bibinfo
  {author} {\bibfnamefont {P.~D.}\ \bibnamefont {Odorico}}, \bibinfo {author}
  {\bibfnamefont {F.}~\bibnamefont {Laio}},\ and\ \bibinfo {author}
  {\bibfnamefont {L.}~\bibnamefont {Ridolfi}},\ }\bibfield  {title} {\bibinfo
  {title} {{A Probabilistic Analysis of Fire-Induced Tree-Grass Coexistence in
  Savannas}},\ }\href {https://doi.org/10.1086/500617} {\bibfield  {journal}
  {\bibinfo  {journal} {The American Naturalist}\ }\textbf {\bibinfo {volume}
  {167}},\ \bibinfo {pages} {78} (\bibinfo {year} {2006})}\BibitemShut
  {NoStop}%
\bibitem [{\citenamefont {Wu}\ and\ \citenamefont {Zhu}(2008)}]{Wu2008a}%
  \BibitemOpen
  \bibfield  {author} {\bibinfo {author} {\bibfnamefont {Y.}~\bibnamefont
  {Wu}}\ and\ \bibinfo {author} {\bibfnamefont {W.~Q.}\ \bibnamefont {Zhu}},\
  }\bibfield  {title} {\bibinfo {title} {{Stochastic analysis of a pulse-type
  prey-predator model}},\ }\href {https://doi.org/10.1103/PhysRevE.77.041911}
  {\bibfield  {journal} {\bibinfo  {journal} {Physical Review E - Statistical,
  Nonlinear, and Soft Matter Physics}\ }\textbf {\bibinfo {volume} {77}},\
  \bibinfo {pages} {1} (\bibinfo {year} {2008})}\BibitemShut {NoStop}%
\bibitem [{\citenamefont {Calabrese}\ \emph {et~al.}(2017)\citenamefont
  {Calabrese}, \citenamefont {Porporato}, \citenamefont {Laio}, \citenamefont
  {Odorico},\ and\ \citenamefont {Ridolfi}}]{Calabrese2017}%
  \BibitemOpen
  \bibfield  {author} {\bibinfo {author} {\bibfnamefont {S.}~\bibnamefont
  {Calabrese}}, \bibinfo {author} {\bibfnamefont {A.}~\bibnamefont
  {Porporato}}, \bibinfo {author} {\bibfnamefont {F.}~\bibnamefont {Laio}},
  \bibinfo {author} {\bibfnamefont {P.~D.}\ \bibnamefont {Odorico}},\ and\
  \bibinfo {author} {\bibfnamefont {L.}~\bibnamefont {Ridolfi}},\ }\bibfield
  {title} {\bibinfo {title} {{Age distribution dynamics with stochastic jumps
  in mortality}},\ }in\ \href {https://doi.org/10.1098/rspa.2017.0451} {\emph
  {\bibinfo {booktitle} {Proceedings of the Royal Society A: Mathematical,
  Physical and Engineering Sciences}}},\ Vol.\ \bibinfo {volume} {473}\
  (\bibinfo {year} {2017})\BibitemShut {NoStop}%
\bibitem [{\citenamefont {Merton}(1976)}]{Merton1976}%
  \BibitemOpen
  \bibfield  {author} {\bibinfo {author} {\bibfnamefont {R.~C.}\ \bibnamefont
  {Merton}},\ }\bibfield  {title} {\bibinfo {title} {{Option pricing when
  underlying stock returns are discontinuous}},\ }\href
  {https://doi.org/10.1016/0304-405X(76)90022-2.} {\bibfield  {journal}
  {\bibinfo  {journal} {Journal of Financial Economics}\ }\textbf {\bibinfo
  {volume} {3}},\ \bibinfo {pages} {125} (\bibinfo {year} {1976})}\BibitemShut
  {NoStop}%
\bibitem [{\citenamefont {Kou}(2002)}]{Kou2002}%
  \BibitemOpen
  \bibfield  {author} {\bibinfo {author} {\bibfnamefont {S.}~\bibnamefont
  {Kou}},\ }\bibfield  {title} {\bibinfo {title} {{A Jump-Diffusion Model for
  Option Pricing}},\ }\href {https://doi.org/10.1287/mnsc.48.8.1086.166}
  {\bibfield  {journal} {\bibinfo  {journal} {Management Science}\ }\textbf
  {\bibinfo {volume} {48}},\ \bibinfo {pages} {1086} (\bibinfo {year}
  {2002})}\BibitemShut {NoStop}%
\bibitem [{\citenamefont {Stidham}(1974)}]{Stidham1974}%
  \BibitemOpen
  \bibfield  {author} {\bibinfo {author} {\bibfnamefont {S.}~\bibnamefont
  {Stidham}},\ }\bibfield  {title} {\bibinfo {title} {{Stochastic clearing
  systems}},\ }\href {https://doi.org/10.1016/0304-4149(74)90014-3} {\bibfield
  {journal} {\bibinfo  {journal} {Stochastic Processes and their Applications}\
  }\textbf {\bibinfo {volume} {2}},\ \bibinfo {pages} {85} (\bibinfo {year}
  {1974})}\BibitemShut {NoStop}%
\bibitem [{\citenamefont {Kella}\ \emph {et~al.}(2003)\citenamefont {Kella},
  \citenamefont {Perry},\ and\ \citenamefont {Stadje}}]{Kella2003}%
  \BibitemOpen
  \bibfield  {author} {\bibinfo {author} {\bibfnamefont {O.}~\bibnamefont
  {Kella}}, \bibinfo {author} {\bibfnamefont {D.}~\bibnamefont {Perry}},\ and\
  \bibinfo {author} {\bibfnamefont {W.}~\bibnamefont {Stadje}},\ }\bibfield
  {title} {\bibinfo {title} {{A stochastic clearing model with a Brownian and a
  compound poisson component}},\ }\href
  {https://doi.org/10.1017/S026996480317101X} {\bibfield  {journal} {\bibinfo
  {journal} {Probability in the Engineering and Informational Sciences}\
  }\textbf {\bibinfo {volume} {17}},\ \bibinfo {pages} {1} (\bibinfo {year}
  {2003})}\BibitemShut {NoStop}%
\bibitem [{\citenamefont {Cox}\ and\ \citenamefont {Miller}(1965)}]{Cox1965}%
  \BibitemOpen
  \bibfield  {author} {\bibinfo {author} {\bibfnamefont {D.~R.}\ \bibnamefont
  {Cox}}\ and\ \bibinfo {author} {\bibfnamefont {H.~D.}\ \bibnamefont
  {Miller}},\ }\href {https://doi.org/10.1201/9780203719152} {\emph {\bibinfo
  {title} {{The Theory of Stochastic Processes}}}}\ (\bibinfo  {publisher}
  {Wiley},\ \bibinfo {address} {New York},\ \bibinfo {year} {1965})\BibitemShut
  {NoStop}%
\bibitem [{\citenamefont {Daly}\ and\ \citenamefont
  {Porporato}(2006)}]{Daly2006}%
  \BibitemOpen
  \bibfield  {author} {\bibinfo {author} {\bibfnamefont {E.}~\bibnamefont
  {Daly}}\ and\ \bibinfo {author} {\bibfnamefont {A.}~\bibnamefont
  {Porporato}},\ }\bibfield  {title} {\bibinfo {title} {{Probabilistic dynamics
  of some jump-diffusion systems}},\ }\href
  {https://doi.org/10.1103/PhysRevE.73.026108} {\bibfield  {journal} {\bibinfo
  {journal} {Physical Review E}\ }\textbf {\bibinfo {volume} {73}},\ \bibinfo
  {pages} {1} (\bibinfo {year} {2006})}\BibitemShut {NoStop}%
\bibitem [{\citenamefont {Daly}\ and\ \citenamefont
  {Porporato}(2010)}]{Daly2010}%
  \BibitemOpen
  \bibfield  {author} {\bibinfo {author} {\bibfnamefont {E.}~\bibnamefont
  {Daly}}\ and\ \bibinfo {author} {\bibfnamefont {A.}~\bibnamefont
  {Porporato}},\ }\bibfield  {title} {\bibinfo {title} {{Effect of different
  jump distributions on the dynamics of jump processes}},\ }\href
  {https://doi.org/10.1103/PhysRevE.81.061133} {\bibfield  {journal} {\bibinfo
  {journal} {Physical Review E}\ }\textbf {\bibinfo {volume} {81}},\ \bibinfo
  {pages} {1} (\bibinfo {year} {2010})}\BibitemShut {NoStop}%
\bibitem [{\citenamefont {Dubkov}\ \emph {et~al.}(2016)\citenamefont {Dubkov},
  \citenamefont {Rudenko},\ and\ \citenamefont {Gurbatov}}]{Dubkov2016}%
  \BibitemOpen
  \bibfield  {author} {\bibinfo {author} {\bibfnamefont {A.~A.}\ \bibnamefont
  {Dubkov}}, \bibinfo {author} {\bibfnamefont {O.~V.}\ \bibnamefont
  {Rudenko}},\ and\ \bibinfo {author} {\bibfnamefont {S.~N.}\ \bibnamefont
  {Gurbatov}},\ }\bibfield  {title} {\bibinfo {title} {{Probability
  characteristics of nonlinear dynamical systems driven by $\delta$-pulse
  noise}},\ }\href {https://doi.org/10.1103/PhysRevE.93.062125} {\bibfield
  {journal} {\bibinfo  {journal} {Physical Review E}\ }\textbf {\bibinfo
  {volume} {93}},\ \bibinfo {pages} {1} (\bibinfo {year} {2016})}\BibitemShut
  {NoStop}%
\bibitem [{\citenamefont {Denisov}\ \emph {et~al.}(2009)\citenamefont
  {Denisov}, \citenamefont {Horsthemke},\ and\ \citenamefont
  {H{\"{a}}nggi}}]{Denisov2009}%
  \BibitemOpen
  \bibfield  {author} {\bibinfo {author} {\bibfnamefont {S.~I.}\ \bibnamefont
  {Denisov}}, \bibinfo {author} {\bibfnamefont {W.}~\bibnamefont
  {Horsthemke}},\ and\ \bibinfo {author} {\bibfnamefont {P.}~\bibnamefont
  {H{\"{a}}nggi}},\ }\bibfield  {title} {\bibinfo {title} {{Generalized
  Fokker-Planck equation: Derivation and exact solutions}},\ }\href
  {https://doi.org/10.1140/epjb/e2009-00126-3} {\bibfield  {journal} {\bibinfo
  {journal} {European Physical Journal B}\ }\textbf {\bibinfo {volume} {68}},\
  \bibinfo {pages} {567} (\bibinfo {year} {2009})},\ \Eprint
  {https://arxiv.org/abs/0808.0274} {arXiv:0808.0274} \BibitemShut {NoStop}%
\bibitem [{\citenamefont {Gardiner}(1983)}]{Gardiner1983}%
  \BibitemOpen
  \bibfield  {author} {\bibinfo {author} {\bibfnamefont {C.~W.}\ \bibnamefont
  {Gardiner}},\ }\href {https://link.springer.com/gp/book/9783540707127} {\emph
  {\bibinfo {title} {{Handbook of Stochastic Methods for Physics, Chemistry and
  the Natural Sciences}}}}\ (\bibinfo  {publisher} {Springer-Verlag},\ \bibinfo
  {address} {Berlin},\ \bibinfo {year} {1983})\BibitemShut {NoStop}%
\bibitem [{\citenamefont {Risken}(1984)}]{Risken1989}%
  \BibitemOpen
  \bibfield  {author} {\bibinfo {author} {\bibfnamefont {H.}~\bibnamefont
  {Risken}},\ }\href {https://doi.org/10.1080/713821438} {\emph {\bibinfo
  {title} {{The Fokker-Planck Equation: Methods of Solution and
  Applications}}}},\ \bibinfo {edition} {2nd}\ ed.\ (\bibinfo  {publisher}
  {Springer-Verlag},\ \bibinfo {address} {Ulm},\ \bibinfo {year}
  {1984})\BibitemShut {NoStop}%
\bibitem [{\citenamefont {Fox}\ \emph {et~al.}(1971)\citenamefont {Fox},
  \citenamefont {Mayers}, \citenamefont {Ockendon},\ and\ \citenamefont
  {Tayler}}]{Fox1971}%
  \BibitemOpen
  \bibfield  {author} {\bibinfo {author} {\bibfnamefont {L.}~\bibnamefont
  {Fox}}, \bibinfo {author} {\bibfnamefont {D.~F.}\ \bibnamefont {Mayers}},
  \bibinfo {author} {\bibfnamefont {J.~R.}\ \bibnamefont {Ockendon}},\ and\
  \bibinfo {author} {\bibfnamefont {A.~B.}\ \bibnamefont {Tayler}},\ }\bibfield
   {title} {\bibinfo {title} {{On a functional differential equation}},\ }\href
  {https://doi.org/10.1134/S1995080217030027} {\bibfield  {journal} {\bibinfo
  {journal} {J. Inst. Maths Applics}\ }\textbf {\bibinfo {volume} {8}},\
  \bibinfo {pages} {271} (\bibinfo {year} {1971})}\BibitemShut {NoStop}%
\bibitem [{\citenamefont {Ockendon}\ and\ \citenamefont
  {Tayler}(1971)}]{Ockendon1971}%
  \BibitemOpen
  \bibfield  {author} {\bibinfo {author} {\bibfnamefont {J.~R.}\ \bibnamefont
  {Ockendon}}\ and\ \bibinfo {author} {\bibfnamefont {A.~B.}\ \bibnamefont
  {Tayler}},\ }\bibfield  {title} {\bibinfo {title} {{The dynamics of a current
  collection system for an electric locomotive}},\ }\href
  {https://doi.org/10.1098/rspa.1971.0078} {\bibfield  {journal} {\bibinfo
  {journal} {Proceedings of the Royal Society of London. A. Mathematical and
  Physical Sciences}\ }\textbf {\bibinfo {volume} {322}},\ \bibinfo {pages}
  {447} (\bibinfo {year} {1971})}\BibitemShut {NoStop}%
\bibitem [{\citenamefont {Hall}\ and\ \citenamefont {Wake}(1989)}]{Hall1989}%
  \BibitemOpen
  \bibfield  {author} {\bibinfo {author} {\bibfnamefont {A.~J.}\ \bibnamefont
  {Hall}}\ and\ \bibinfo {author} {\bibfnamefont {G.~C.}\ \bibnamefont
  {Wake}},\ }\bibfield  {title} {\bibinfo {title} {{A functional differential
  equation arising in modelling of cell growth}},\ }\href
  {https://doi.org/10.1017/s0334270000006366} {\bibfield  {journal} {\bibinfo
  {journal} {The Journal of the Australian Mathematical Society. Series B.
  Applied Mathematics}\ }\textbf {\bibinfo {volume} {30}},\ \bibinfo {pages}
  {424} (\bibinfo {year} {1989})}\BibitemShut {NoStop}%
\bibitem [{\citenamefont {van Brunt}\ \emph {et~al.}(2018)\citenamefont {van
  Brunt}, \citenamefont {Zaidi},\ and\ \citenamefont {Lynch}}]{VanBrunt2018}%
  \BibitemOpen
  \bibfield  {author} {\bibinfo {author} {\bibfnamefont {B.}~\bibnamefont {van
  Brunt}}, \bibinfo {author} {\bibfnamefont {A.~A.}\ \bibnamefont {Zaidi}},\
  and\ \bibinfo {author} {\bibfnamefont {T.}~\bibnamefont {Lynch}},\ }\bibfield
   {title} {\bibinfo {title} {{Cell Division And The Pantograph Equation}},\
  }\href {https://doi.org/10.1051/proc/201862158} {\bibfield  {journal}
  {\bibinfo  {journal} {ESAIM: Proceedings and Surveys}\ }\textbf {\bibinfo
  {volume} {62}},\ \bibinfo {pages} {158} (\bibinfo {year} {2018})}\BibitemShut
  {NoStop}%
\bibitem [{\citenamefont {Efendiev}\ \emph {et~al.}(2018)\citenamefont
  {Efendiev}, \citenamefont {van Brunt}, \citenamefont {Wake},\ and\
  \citenamefont {Zaidi}}]{Efendiev2018}%
  \BibitemOpen
  \bibfield  {author} {\bibinfo {author} {\bibfnamefont {M.}~\bibnamefont
  {Efendiev}}, \bibinfo {author} {\bibfnamefont {B.}~\bibnamefont
  {van Brunt}}, \bibinfo {author} {\bibfnamefont {G.~C.}\ \bibnamefont
  {Wake}},\ and\ \bibinfo {author} {\bibfnamefont {A.~A.}\ \bibnamefont
  {Zaidi}},\ }\bibfield  {title} {\bibinfo {title} {{A functional partial
  differential equation arising in a cell growth model with dispersion}},\
  }\href {https://doi.org/10.1002/mma.4684} {\bibfield  {journal} {\bibinfo
  {journal} {Mathematical Methods in the Applied Sciences}\ }\textbf {\bibinfo
  {volume} {41}},\ \bibinfo {pages} {1541} (\bibinfo {year}
  {2018})}\BibitemShut {NoStop}%
\bibitem [{\citenamefont {Zaidi}\ \emph {et~al.}(2015)\citenamefont {Zaidi},
  \citenamefont {{Van Brunt}},\ and\ \citenamefont {Wake}}]{Zaidi2015}%
  \BibitemOpen
  \bibfield  {author} {\bibinfo {author} {\bibfnamefont {A.~A.}\ \bibnamefont
  {Zaidi}}, \bibinfo {author} {\bibfnamefont {B.}~\bibnamefont {{Van Brunt}}},\
  and\ \bibinfo {author} {\bibfnamefont {G.~C.}\ \bibnamefont {Wake}},\
  }\bibfield  {title} {\bibinfo {title} {{Solutions to an advanced functional
  partial differential equation of the pantograph type}},\ }\bibfield
  {journal} {\bibinfo  {journal} {Proc. Roy. Soc. London, Series A.}\ }\textbf
  {\bibinfo {volume} {471}},\ \href {https://doi.org/10.1098/rspa.2014.0947}
  {10.1098/rspa.2014.0947} (\bibinfo {year} {2015})\BibitemShut {NoStop}%
\bibitem [{\citenamefont {Zaidi}\ and\ \citenamefont {{Van
  Brunt}}(2021)}]{Zaidi2021}%
  \BibitemOpen
  \bibfield  {author} {\bibinfo {author} {\bibfnamefont {A.~A.}\ \bibnamefont
  {Zaidi}}\ and\ \bibinfo {author} {\bibfnamefont {B.}~\bibnamefont {{Van
  Brunt}}},\ }\bibfield  {title} {\bibinfo {title} {{Asymmetrical cell division
  with exponential growth}},\ }\href
  {https://doi.org/10.1017/S1446181121000109} {\bibfield  {journal} {\bibinfo
  {journal} {ANZIAM Journal}\ }\textbf {\bibinfo {volume} {63}},\ \bibinfo
  {pages} {70} (\bibinfo {year} {2021})}\BibitemShut {NoStop}%
\bibitem [{\citenamefont {Barik}\ \emph {et~al.}(2006)\citenamefont {Barik},
  \citenamefont {Ghosh},\ and\ \citenamefont {Ray}}]{Barik2006}%
  \BibitemOpen
  \bibfield  {author} {\bibinfo {author} {\bibfnamefont {D.}~\bibnamefont
  {Barik}}, \bibinfo {author} {\bibfnamefont {P.~K.}\ \bibnamefont {Ghosh}},\
  and\ \bibinfo {author} {\bibfnamefont {D.~S.}\ \bibnamefont {Ray}},\
  }\bibfield  {title} {\bibinfo {title} {{Langevin dynamics with dichotomous
  noise; Direct simulation and applications}},\ }\bibfield  {journal} {\bibinfo
   {journal} {Journal of Statistical Mechanics: Theory and Experiment}\ }\href
  {https://doi.org/10.1088/1742-5468/2006/03/P03010}
  {10.1088/1742-5468/2006/03/P03010} (\bibinfo {year} {2006}),\ \Eprint
  {https://arxiv.org/abs/0602556} {arXiv:0602556 [cond-mat]} \BibitemShut
  {NoStop}%
\bibitem [{\citenamefont {Kim}(1998)}]{Kim1998}%
  \BibitemOpen
  \bibfield  {author} {\bibinfo {author} {\bibfnamefont {H.~K.}\ \bibnamefont
  {Kim}},\ }\emph {\bibinfo {title} {{Advanced second order functional
  differential equations}}},\ \href
  {https://mro.massey.ac.nz/handle/10179/2393} {Ph.D. thesis},\ \bibinfo
  {school} {Massey University} (\bibinfo {year} {1998})\BibitemShut {NoStop}%
\bibitem [{\citenamefont {{Van Brunt}}\ and\ \citenamefont
  {Vlieg-Hulstman}(2011)}]{VanBrunt2011}%
  \BibitemOpen
  \bibfield  {author} {\bibinfo {author} {\bibfnamefont {B.}~\bibnamefont {{Van
  Brunt}}}\ and\ \bibinfo {author} {\bibfnamefont {M.}~\bibnamefont
  {Vlieg-Hulstman}},\ }\bibfield  {title} {\bibinfo {title} {{Eigenfunctions
  arising from a first-order functional differential equation in a cell growth
  model}},\ }\href {https://doi.org/10.1017/S1446181111000575} {\bibfield
  {journal} {\bibinfo  {journal} {ANZIAM Journal}\ }\textbf {\bibinfo {volume}
  {52}},\ \bibinfo {pages} {46} (\bibinfo {year} {2011})}\BibitemShut {NoStop}%
\bibitem [{\citenamefont {Masoliver}(2018)}]{Masoliver2018}%
  \BibitemOpen
  \bibfield  {author} {\bibinfo {author} {\bibfnamefont {J.}~\bibnamefont
  {Masoliver}},\ }\href {https://doi.org/10.1142/10578} {\emph {\bibinfo
  {title} {Random Processes: First Passage and Escape}}}\ (\bibinfo
  {publisher} {World Scientific},\ \bibinfo {year} {2018})\BibitemShut
  {NoStop}%
\bibitem [{\citenamefont {Balakrishnan}(2021)}]{Balakrishnan2021}%
  \BibitemOpen
  \bibfield  {author} {\bibinfo {author} {\bibfnamefont {V.}~\bibnamefont
  {Balakrishnan}},\ }\href {https://doi.org/10.1007/978-3-030-62233-6} {\emph
  {\bibinfo {title} {Elements of Nonequilibrium Statistical Mechanics}}}\
  (\bibinfo  {publisher} {Springer},\ \bibinfo {address} {Berlin},\ \bibinfo
  {year} {2021})\BibitemShut {NoStop}%
\bibitem [{\citenamefont {Andrews}\ \emph {et~al.}(1999)\citenamefont
  {Andrews}, \citenamefont {Askey},\ and\ \citenamefont {Roy}}]{Andrews1999}%
  \BibitemOpen
  \bibfield  {author} {\bibinfo {author} {\bibfnamefont {G.}~\bibnamefont
  {Andrews}}, \bibinfo {author} {\bibfnamefont {R.}~\bibnamefont {Askey}},\
  and\ \bibinfo {author} {\bibfnamefont {R.}~\bibnamefont {Roy}},\ }\href
  {https://doi.org/10.1017/CBO9781107325937} {\emph {\bibinfo {title} {{Special
  Functions}}}}\ (\bibinfo  {publisher} {Cambridge University Press},\ \bibinfo
  {address} {Cambridge},\ \bibinfo {year} {1999})\BibitemShut {NoStop}%
\end{thebibliography}%


\end{document}